# Negative Refractive Index in Optics of Metal-Dielectric Composites

**A. V. Kildishev , W. Cai, U. K. Chettiar, H.-K. Yuan, A. K. Sarychev*, V. P. Drachev, and V. M. Shalaev**

School of Electrical and Computer Engineering, 465 Northwestern Ave., W. Lafayette, IN 47907

* Ethertronics Inc., 9605 Scranton Road, San Diego, CA 92121

**Abstract**: Specially designed metal-dielectric composites can have a negative refractive index in the optical range. Specifically, it is shown that arrays of single and paired nanorods can provide such negative refraction. For pairs of metal rods, a negative refractive index has been observed at 1.5 μm. The inverted structure of paired voids in metal films may also exhibit a negative refractive index. A similar effect can be accomplished with metal strips in which the refractive index can reach −2. The refractive index retrieval procedure and the critical role of light phases in determining the refractive index is discussed.

# Introduction

In the recent few years, there has been a strong interest in novel optical media which have become known as left-handed materials (LHMs) or negative index materials (NIMs). Such materials have not been discovered as natural substances or crystals, but rather are artificial, man-made materials. The optical properties of such media were considered in early papers by two Russian physicists, Mandel'shtam[1] and Veselago,[2] although much earlier works on negative phase velocity and its consequences belong to Lamb[3] (in hydrodynamics) and Schuster[4] (in optics).

In NIMs, $\vec{k}$, $\vec{E}$, and $\vec{H}$ form a left-handed set of vectors, and were therefore named LHMs by Veselago. As a result of the negative index of refraction and negative phase velocity, these artificial materials exhibit a number of extraordinary features, including an inverse Snell's law relationship, a reversed Doppler shift, and reversed Cherenkov radiation. These features suggest a flexible regulation of light propagation in these media and facilitate new, fascinating applications.

The most recent successful efforts to demonstrate negative refraction have been inspired by Pendry's revision[5] of the Veselago lens, which renewed interest in the practical aspects of the earlier papers. Pendry predicted that a NIM lens can act as a "superlens" by providing spatial resolution beyond the diffraction limit. Fig. 1 compares the simplest case of refraction at a single interface between vacuum and common, positive refractive index material (Fig. 1(a)) versus refraction at the interface with a NIM (Fig. 1(b)). For any oblique angle of incidence $\theta_i$, the tangential wavevector $\vec{k}_\parallel$ of an incident plane wave from the vacuum side must remain continuous across the interface. This is the case for both positive and negative refractive indicies. However, in contrast to a normal, positive refraction material as the light passes from vacuum into a NIM, the wavevector component normal to the interface ($\vec{k}_\perp$) must change the sign. As a result, the total refracted wavevector is on the same side of the normal as the incident wavevector. Figure 1(b) illustrates this effect as the reversed Snell's law, where the angles of reflection ($\theta_r$) and transmission ($\theta_t$) are reversed ($\theta_r = \theta_t$) for the case of $n = -1$. This reversal suggests insightful consequences for the imaging applications of NIMs, the most critical of which is the 'perfect lens' as illustrated in Fig. 1(c). For refracted rays from a point source, a planar NIM slab of sufficient thickness with $n = -1$ should first focus the rays inside the NIM and then re-focus them again behind the slab. At the perfect lensing condition of $n = -1$, this focusing property is extraordinary, and the resolution limit intrinsic to conventional imaging no longer applies to imaging with a NIM slab as shown by Pendry.[5] The essence of the effect is that a NIM compensates the usual decay of the evanescent waves. Contrary to a conventional imaging device, in a super lens these evanescent waves are recovered by the NIM and the image is perfectly reconstructed.[5] The perfect lensing requirements ($\mathrm{Re}(n) = -1$, $\mathrm{Im}(n) = 0$) are difficult conditions to meet. The first requirement means that *n* is wavelength-dependent and the perfect lens is restricted to work at a single wavelength. The second requirement implies that there is no absorption in the NIM. In reality, losses are always present in the NIM and can dramatically diminish the resolution.[6,7]

The development of optical NIMs is closely connected to studies of periodic arrays of elementary scatterers, where, for example, frequency selective surfaces (FSS) arranged from those arrays have been used as narrowband filters for plane waves. The major resonant elements of optical metal-dielectric composites (metallic spheres, disks, rods, and their inversions, i.e. circular and elliptic holes) have been inherited from earlier FSS designs, together with evident orthotropic properties and angular dependences. Another source of earlier expertise is found in the studies of artificial dielectrics (ADs), where the homogenization method and approaches to define equivalent electromagnetic material properties have been examined under resonant conditions. Predictions of anomalously large effective permeability and permittivity due to resonant regimes of elementary scatterers have also been made for ADs arranged from periodic metal-dielectric structures[8].

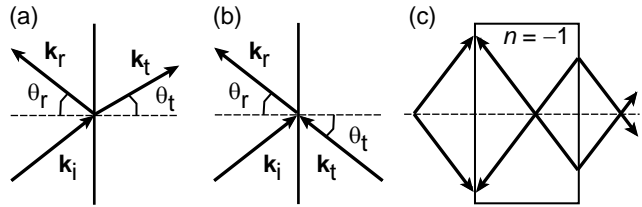

**Fig. 1. (a) Snell's law at an interface between vacuum and a positive refractive index material; (b) reversed Snell's law at the interface with NIM (n = −1); and (c) the superlens.**

While negative permittivity in the optical range is easy to attain for metals, there is no magnetic response for naturally-occurring materials at such high frequencies. Recent theories and experiments showed that a magnetic response and negative permeability can be accomplished in the terahertz spectral ranges by using parallel rods, split-ring resonators (SRRs) and other structures.[9-14] As predicted[13] and experimentally demonstrated,[14] u-shaped structures are particularly well suited for magnetism at optical frequencies. Recently, light propagation through an interface that mimics negative refraction has also been found in two-dimensional photonic crystals (PCs).[15] It has been shown that under certain conditions, unique focusing effects in PCs are also possible.[16]

Up to the microwave frequencies the fabrication of metal-dielectric composites can follow practically any pattern suggested either by human intuition or computer-aided tools with evolutionary optimization. Yet at optical frequencies, the best possible design of metal-dielectric composite NIM should overcome two substantial difficulties: severe fabrication constrains and increased losses. For these reasons, this paper addresses the simplest geometries of resonant 3D structures. Such structures include coupled metal rods in a dielectric host[17-21] (fabricated, for example, through electron-beam lithography (EBL)) and inversions of rods, i.e. coupled dielectric holes of elliptical (spherical,[22] in the limiting case) or rectangular shape[23] (fabricated, for example, by etching tri-layer metal-dielectric-metal films by ion beam etching (IBE) or through interferometric lithography).

The first experimental realization of a negative refractive index in the optical range (at 1.5 µm) was accomplished with paired metal nanorods in a dielectric,[21] and then for the inverted system of paired dielectric voids in a metal.[22,23] We note that inverted NIMs, i.e. elliptical or rectangular dielectric voids in metal films,[23] are physically equivalent to paired metal rods in a dielectric host, in accordance with the Babinet principle.[24] We also note here Ref. [25] that considers other interesting optical properties of metal rods, although these properties are not related directly to NIMs.

In the current paper, in addition to paired metal rods and dielectric voids, we also study two-dimensional coupled-strip composites as a basis for further comprehensive studies of NIMs.

## Homogenized optical parameters of NIMs

*Equivalent multilayer structure at normal incidence*

Our NIM designs are confined in a thin layer placed on a thick substrate; therefore, a direct measurement can not resolve the reversed refraction due to insufficient optical length. Fortunately, accurate indirect measurements work very well to retrieve the effective optical constants of the NIM.

First, we ascribe an effective refractive index to a layer of the NIM as if it were a layer of homogeneous medium. This assumption suggests that the periodic structure of NIM does not diffract the incident plane wave. Then we consider a straightforward direct problem of plane wave propagation through a multilayer structure of homogeneous materials at normal incidence, as shown in Fig. 2. The electric field at the initial interface on the source side ($E_i^- = E_{i,1}$) is first compared to the field transmitted through the same boundary ($E^- = E_{t,1}$) and then compared to the transmitted field ($E^+ = E_{t,\nu_{\max}}$) at the last interface of the back side. Then, provided that the multilayer structure is surrounded by vacuum (i.e., $n_0 = n_{\nu_{\max}} = 1$), the complex reflection coefficient $r = E^-/E_i - 1$ and the transmission coefficient $t = E^+/E_i$ of the entire structure can be obtained from the following equations:

$$t = \frac{2}{z_{11} + z_{12} + z_{21} + z_{22}}, \quad (1)$$

$$r = \frac{(z_{11} + z_{12}) - (z_{21} + z_{22})}{z_{11} + z_{12} + z_{21} + z_{22}}, \quad (2)$$

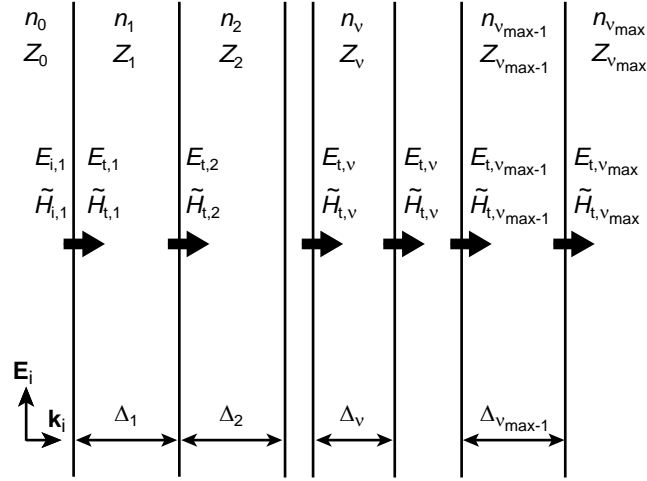

**Fig. 2.** A multilayer structure illuminated from left to right by a monochromatic plane wave at normal incidence. Each layer is made of a homogeneous material and characterized by refractive index (*n*), impedance (*Z*) and thickness (Δ).

Here $z_{11}$, $z_{12}$, $z_{21}$, and $z_{22}$ are the elements of the total characteristic matrix ($\mathbf{Z}$) of a given multilayer structure obtained as the matrix product of the individual characteristic matrices of each layer, i.e.

$$\mathbf{Z} = \prod_{\nu=1}^{\max(\nu)} \mathbf{Z}_\nu .\qquad(3)$$

Provided that for each layer the relative bulk material properties ($\varepsilon_\nu$ and $\mu_\nu$) and therefore the refractive index ($n_\nu = \sqrt{\varepsilon_\nu \mu_\nu}$) and intrinsic impedance ($Z_\nu = \sqrt{\mu_\nu/\varepsilon_\nu}$) are known, the characteristic matrix of the given layer is defined as

$$\mathbf{Z}_\nu = \begin{pmatrix} \cos(n_\nu k \Delta_\nu) & -\iota Z_\nu \sin(n_\nu k \Delta_\nu) \\ -\iota Z_\nu^{-1} \sin(n_\nu k \Delta_\nu) & \cos(n_\nu k \Delta_\nu) \end{pmatrix},\qquad(4)$$

where $\Delta_\nu$ is the thickness of $\nu$-th layer. As a result, similar to Ref. [26], for a single layer in vacuum, we have a remarkably symmetric pair of equations:

$$\cosh \zeta = \left(1 - t^2 + r^2\right)/(2r), \qquad (6)$$

$$\cos(nk\Delta) = \left(1 - r^2 + t^2\right)/(2t), \qquad (7)$$

where $\zeta \triangleq 2\coth^{-1} Z$.

If the refractive indices of a source-side medium and a back-side medium differ from that of vacuum, then using normalized values of the magnetic field intensity $\tilde{H} = \left(\mu_0/\varepsilon_0\right)^{1/2} H$ and defining the electric transmission coefficient ($t_{E,\nu} = E_{t,\nu}/E_{i,1}$) in addition to the magnetic transmission coefficient ($t_{H,\nu} = \tilde{H}_{t,\nu}/E_{i,1}$), we arrive at a matrix identity for a given layer:

$$\begin{pmatrix} t_{E,\nu} \\ t_{H,\nu} \end{pmatrix} = \mathbf{Z}_\nu \begin{pmatrix} t_{E,\nu+1} \\ t_{H,\nu+1} \end{pmatrix}. \qquad (8)$$

In our case, the only unknown parameters in the entire multilayer structure are those of the NIM ($n_{NIM} = n_\nu$ and $Z_{NIM} = Z_\nu$), and the above equation can be inverted to restore the parameters through a set of equations

$$Z_\nu^2 = \left(t_{E,\nu}^2 - t_{E,\nu+1}^2\right)/\left(t_{H,\nu}^2 - t_{H,\nu+1}^2\right), \qquad (9)$$

and

$$n_\nu = N_\nu / (k\Delta_\nu), \tag{10}$$

where

$$N_\nu = \cos^{-1}\left(\frac{t_{E,\nu}t_{H,\nu} + t_{E,\nu+1}t_{H,\nu+1}}{t_{E,\nu}t_{H,\nu+1} + t_{E,\nu+1}t_{H,\nu}}\right). \tag{11}$$

We assume that the coefficients $t_E$ and $t_H$ are known from both sides of the layer, either from calculations or through measurements.

## Restoration of the refractive index

Although, Eqs. (9) and (10) seem quite straightforward, physically sound restrictions should be applied for the both expressions.[26] Since the material of the $\nu$-th layer is passive, we choose appropriate signs in Eqs. (9) and (10) in order to obey the restrictions $\mathrm{Re}(Z) > 0$ and $\mathrm{Im}(n) > 0$. Thus, the refractive index $n_\nu = n'_\nu + \iota n''_\nu$ is given by:

$$n'_\nu = \left(sign(N''_\nu)N'_\nu + 2\pi l\right)/(k\Delta_\nu), \tag{12}$$

$$n''_\nu = sign(N''_\nu)\, N''_\nu / (k\Delta_\nu), \tag{13}$$

where $sign(x)$ is equal to 1 if $x \geq 0$, and to $-1$ otherwise; $N'_\nu = \mathrm{Re}(N_\nu)$ and $N''_\nu = \mathrm{Im}(N_\nu)$.

Since $n'$ has multiple branches, to avoid ambiguities in selecting a phase-adjusting integer $l$ in Eq. (12) one should start the restoration of $n'_\nu$ from a higher wavelength (far away from resonances) and obtain physically sound values of $n'_\nu$. Then, the wavelength should be moved toward shorter values while simultaneously adjusting the values of $l$ in Eq. (12) to obtain a continuous behavior for $n'_\nu$. To achieve accurate and unambiguous results, the restored layer should be much thinner than (at least) the longest wavelength in the sweep range and the sweep step size should be adaptively decreased at resonances to provide an adequate number of points in steep segments.

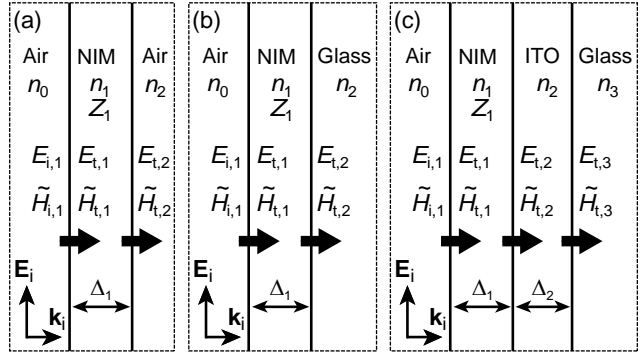

**Fig. 3. (a) A NIM layer in air ($n_0 = n_2 = 1$); (b) a single NIM layer on bare glass substrate ($n_0 = 1$, $n_2 = n_s$); and (c) a NIM layer on ITO-glass substrate ($n_0 = 1$, $n_2 = n_{ITO}$, $n_3 = n_s$).**

In Eqs. (9) – (13), the coefficients $t_{E,\nu}$, $t_{E,\nu+1}$ and $t_{H,\nu}$, $t_{H,\nu+1}$ are considered to be known at both sides of the layer. These coefficients can be recursively restored through Eq. (8) by either direct or back propagation.

Consider, for example, the most frequent cases shown in Fig. 3, where diagram (a) depicts a NIM layer in air ($n_0 = n_2 = 1$); (b) shows a NIM layer on thick glass ($n_0 = 1$, $n_2 = n_s$, with $n_s$ being the refractive index of glass); and (c) shows a NIM layer on an ITO-coated glass substrate ($n_0 = 1$, $n_3 = n_s$, and $n_2 = n_{ITO}$). The refractive index of a *single NIM layer in air*[26] with

$$t_{E,1} = 1 + r, \; t_{H,1} = 1 - r, \tag{14}$$

and $t_{E,2} = t_{H,2} = t$ is already given by Eq. (6) and (7).

The NIM-air interface coefficients $t_{E,1}$ and $t_{H,1}$ of a *NIM layer on a thick bare glass substrate* (Fig. 3(b)) are also defined by Eq. (14), but the coefficients at the NIM-substrate interface are given by $t_{E,2} = t$ and $t_{H,2} = n_s t$, i.e.

$$N_1 = \cos^{-1}\left[\frac{1 - r^2 + n_s t^2}{(n_s + 1)t + rt(n_s - 1)}\right], \quad (15)$$

where $t$ is calculated (measured) in the substrate.

Coefficients $t_{E,1}$ and $t_{H,1}$ at the NIM-air interface of a *NIM layer on an ITO-glass substrate* as in Fig. 3(c) are also defined by Eq. (14), but back side coefficients should be first calculated (measured) in the substrate at the ITO-glass interface ($t_{E,3} = t$, $t_{H,3} = n_s t$) and then back-propagated using Eq. (8), i.e.

$$\begin{pmatrix} t_{E,2} \\ t_{H,2} \end{pmatrix} = \mathbf{Z}_2 \begin{pmatrix} t \\ n_s t \end{pmatrix}, \quad (16)$$

where $\mathbf{Z}_2 = \begin{pmatrix} \cos(n_2 k \Delta_2) & -\iota n_2^{-1} \sin(n_2 k \Delta_2) \\ -\iota n_2 \sin(n_2 k \Delta_2) & \cos(n_2 k \Delta_2) \end{pmatrix}$ with $n_2 = n_{ITO}$. For all cases in Fig. 3, the values of $n_1'$ and $n_1''$ are finally calculated using Eqs. (11)–(13) for $\nu = 1$.

## *Phase-based approximations of the refractive index*

Consider the simple NIM structure depicted in Fig. 4(a). The NIM consists of a periodic 2D array of identical gold strips separated by a silica spacer. The array is periodic in the vertical direction and its simple geometry is characterized by a period $h$, a gold strip width $w$, a strip thickness $\delta$, and a total NIM layer thickness $\Delta_1$. Fig. 4(b) shows the transverse magnetic field $H_z$ between the strips calculated using the finite element method (FEM) for two samples with different gold thickness (13 nm and 19 nm). All other dimensions in both case are identical ($h =$ 480 nm, $w =$ 450 nm, and $\Delta_1 =$ 160 nm). The values of $H_z$ in both field maps of Fig. 4(b) are normalized by the instantaneous magnitude of the incident field, which is adjusted to arrive at a positive maximum half way between the strips. The magnetic field in both cases is calculated at a minimal refractive index (i.e., at λ ≈ 850 nm for $\delta$ = 13 nm and λ ≈ 900 nm for $\delta$ = 19 nm) and is normalized using the maximal incident field. Figs. 4(c) and (d) depict the reflectance $R$, absorbance $A$, and the refractive index $n'$ of the strips as functions of the wavelength of incident light.

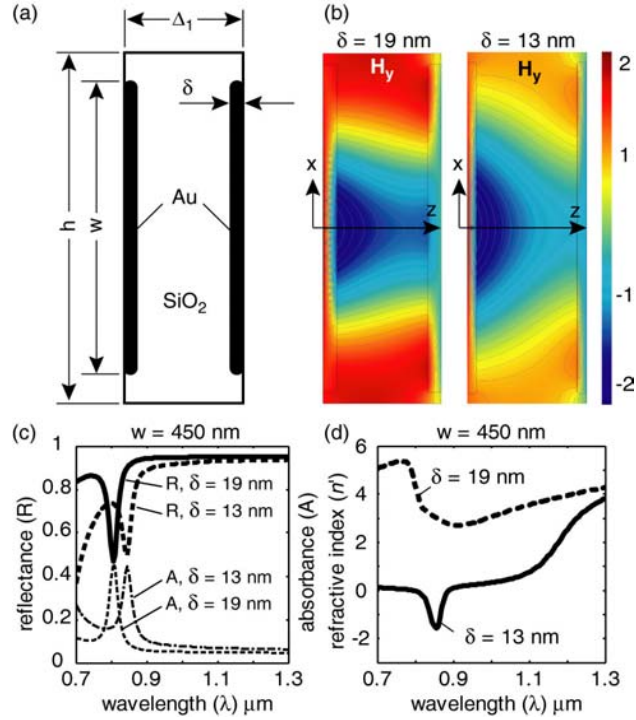

**Fig. 4. (a) A cross-section view of a 2D NIM layer arranged from coupled gold strips separated by a layer of silica. (b) Transverse magnetic field $H_z$ between the strips for two different gold thicknesses for vertical polarization of the incident field. (c) Reflectance R, absorbance A, and (d) the refractive index n′ of the strips vs. the wavelength of the incident light, (c) and (d).**

The simulations indicate that the index of refraction for the structure in Fig. 4(a) depends dramatically on thickness $\delta$. Notice how a six nm increase in thickness (about 30 %) weakens the effective diamagnetic properties of the structure (Fig. 4(b)) and cancels the negative refraction effect in Fig. 4(d), simultaneously shifting the resonances of $A$ and $R$ toward shorter wavelengths in Fig. 4(c). The quantitative changes in absorbed and reflected energy are to be expected, and an analysis in terms of reflectivity and transmission is sometimes used to restore the refractive index[22]. Unfortunately these quantitative differences can be easily distorted by experiment-simulation mismatches and measurement errors. The use of only the magnitude changes of reflectivity and transmission complicates the examination of the resonant behavior of $n$, which for the most part follows phase changes in the transmitted and reflected light. Fig. 4 clearly shows that structures with very similar magnitudes of reflectance (transmittance) may have dramatically different refractive indices, which emphases the role of phase measurements in finding the refraction.

To show just how indicative the phase changes can be to the negative refraction behavior we consider the following two phase-based approximations for $n'$,

$$n' \approx \frac{\arg t}{k\Delta} = \frac{\tau}{k\Delta}, (|r| \ll 1), \qquad (17)$$

$$n' \approx \psi = \frac{\left(\arg t - \arg r - \dfrac{\pi}{2}\right)}{k\Delta}, \ (|r| \to 1), \qquad (18)$$

which are obtained from Eq. (7) for either low or large reflectance. For example, taking Eq. (7) at the limit of $|r| \ll 1$ and using $t = \exp[\iota(\tau - \iota \ln|t|)]$, we arrive at Eq. (17), and the approximation for the imaginary part follows as $n'' \approx \ln|t|/(k\Delta)$. It is interesting to note that according to Eqs. (17) and (18), the refractive index is fully determined by phases only, in the corresponding limiting cases.

To provide a test example for Eqs. (17) and (18), we approximate the values of refractive index obtained from FEM simulations of the 2D structure of Fig. 4(a). Fig. 5 depicts the refractive index of periodic paired strips retrieved from the exact retrieval formula of Eq. (7) and independently obtained from the approximate formulas of Eqs. (17) and (18) for six different geometries. First, note that the negative refractive index can reach large magnitudes for such structures close to -2 (see Fig. 5e and 5f). Common parameters for all cases are the period ($h =$ 480 nm), and the total layer thickness ($\Delta =$ 140 nm). The strips in (a), (c), and (e) are 440 nm wide and 16, 15, and 14 nm thick, respectively. The 450 nm strips in (b), (d), and (f) are 17, 16, and 15 nm thick, respectively. To simulate the complex permittivity of gold in FEM models, the Drude model is used with parameters selected to match the experimental optical constants of bulk gold. Each simulation begins from a wavelength of 2.5 μm (not shown) and then continues toward shorter values with a gradual step size variation from 50 to 1 nm at steep segments. The cases of Fig. 5 are specifically selected to represent the main possible scenarios, i.e. positive refraction, shown in (a) and (b); negative refraction, shown in (e) and (f); or a transition to $n' < 1$, shown in (c) and (d).

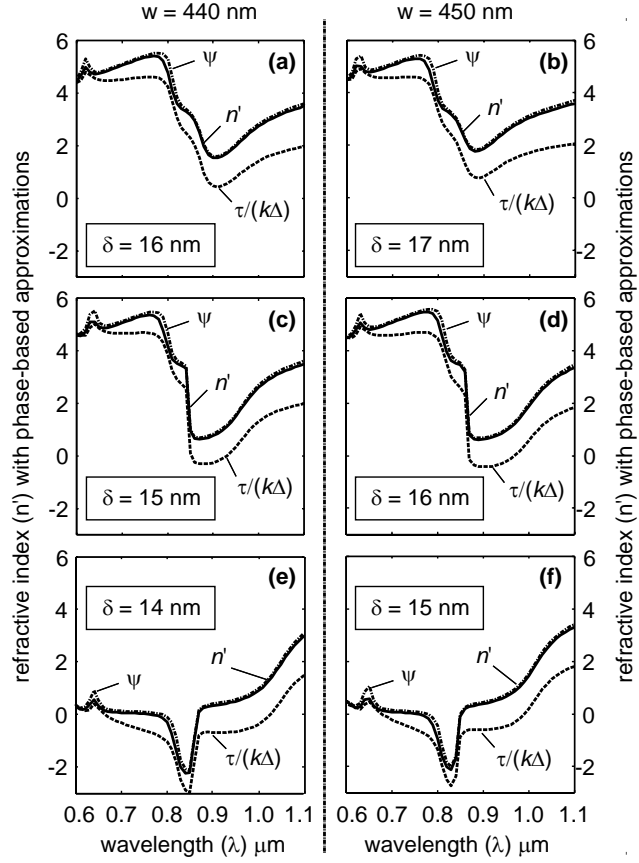

**Fig. 5. The refractive index of periodic paired strips obtained for two different geometries from FEM simulations and approximations at the same wavelength range. Strips in (a), (c), and (e) are 440 nm wide, while those in (b), (d), and (f) are 450 nm wide. Approximating function ψ is given by Eq. (18).**

Notice that Eqs. (17) and (18) do not include any magnitudes and suggest an instrumental role for phase differences $\tau = \arg t$ and $\rho = \arg r$ in representing NIM features. Indeed, in all cases both approximations illustrate well the changes of $n'$. Certainly, the approximate formula of Eq. (18) works better than the simpler formula of Eq. (17). This is because the reflection is quite large at short waves, and it is increasing toward longer wavelengths, as shown in Fig. 4(c).

## High-precision phase measurements of thin NIMs

We performed phase measurements using polarization and walk-off interferometry schemes depicted in Fig. 6 for transmitted light phases ($\tau$). The phase differences in reflection ($\rho$) are measured in a similar manner. In both schematic diagrams LC is a liquid crystal phase compensator, PD is a photo-detector, and P is a linear polarizer (the axes of the input/output polarizers are parallel at 45°).

In the polarization interferometer shown in Fig. 6(a), the two optical channels have a common geometrical path and differ only by the polarization of light. The phase differences caused by anisotropy of a refractive material in transmission $\Delta\tau$ (or reflection $\Delta\rho$) are measured between orthogonally-polarized waves $\Delta\tau = \tau_{\parallel} - \tau_{\perp}$ (or $\Delta\rho = \rho_{\parallel} - \rho_{\perp}$). Notice that the phase acquired in the substrate contributes nothing to either $\Delta\tau$ or $\Delta\rho$.

The walk-off interferometer, shown in Fig. 6(b), has two optical channels which differ in geometrical path; this yields a phase shift introduced by a NIM sample in transmission ($\tau_s$) or reflection ($\rho_s$) relative to a reference ($\tau_{air}$ or $\rho_{air}$) so that $\delta\tau = \tau_s - \tau_{air}$ or $\delta\rho = \rho_s - \rho_{air}$. A layer of air with the same thickness as the NIM layer is used as the reference. Both the reference and sample beams go through the substrate so that the phase acquired in the substrate does not contribute to the measured phase shift $\delta\tau$ and $\delta\rho$, provided that the substrate has no deviations in optical thickness. The walk-off effect in anisotropic crystals (ACs) is employed to separate the two beams and then bring them together to produce interference. The phase shifts $\delta\tau_{\parallel}$ ($\delta\rho_{\parallel}$) and $\delta\tau_{\perp}$ ($\delta\rho_{\perp}$) are measured for two light polarizations using a set of diode lasers and a tunable erbium laser, and their difference is compared with the phase anisotropy $\Delta \tau$ ($\Delta \rho$) obtained from polarization interferometry, since $\Delta\tau = \delta\tau_\parallel - \delta\tau_\perp$ and $\Delta\rho = \delta\rho_\parallel - \delta\rho_\perp$.

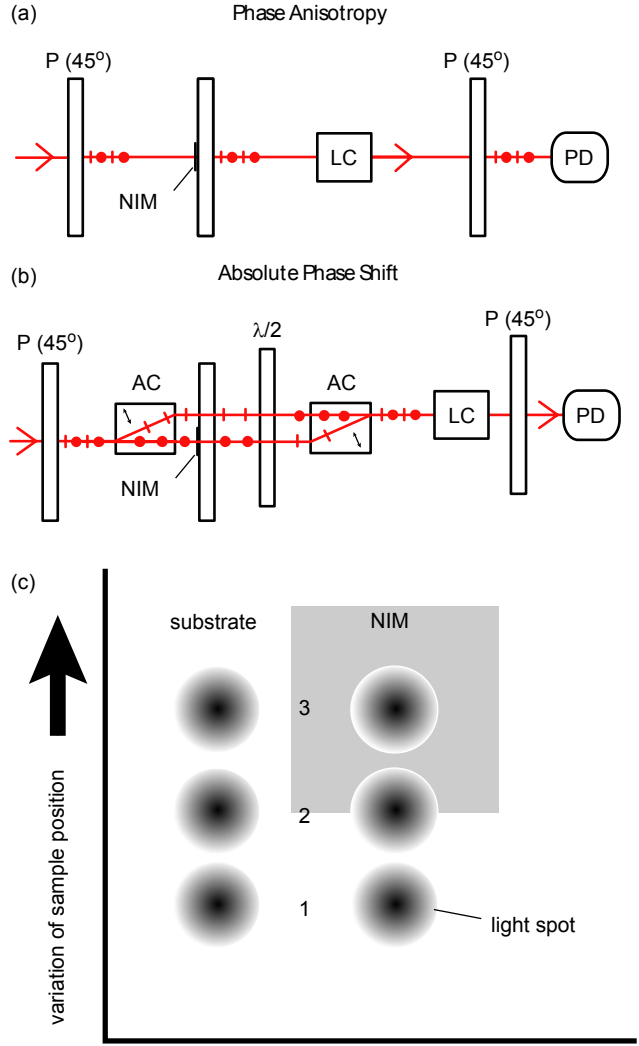

**Fig. 6. Schematic of polarization (a) and walk-off (b) interferometers for measuring phase anisotropy and absolute phase induced by a NIM sample. LC is a liquid crystal phase compensator, P is a 45-degree linear polarizer, AC is an anisotropic calcite crystal with walk-off effect, λ/2 is a half-wave plate and PD is a photo-detector. (c) Calibration of the walk-off interferometry using a gradual change in the spot position relative to the sample: (1) substrate-only position, (2) intermediate position, and (3) final position at the center of NIM sample.**

The instrumental error of the phase anisotropy measurement by polarization interferometer is ±1.7°. We note that variations in the substrate thickness do not affect the results of our phase anisotropy measurements, which is typical for common path interferometers. In the case of the walk-off interferometer, the thickness variation gives an additional source of error, causing the error for the absolute phase shift measurements to increase up to ±4°.

We note here two important details of the measurements. First, by varying the lateral position of the sample relative to the optical beam, the phase shift is smoothly changed from zero when the beam is outside the sample (position 1 in Fig. 6(c)), to the measured value when the beam is at the sample center (position 3). The introduction of intermediate points, such as position 2, eliminates the $2\pi$ uncertainty in measurement. Secondly, the linear polarization of light is well preserved after propagation through the sample for both light polarizations (parallel and perpendicular to the rods). Specifically, the light ellipticity (the intensity ratio for the two components) changes only from $10^{-3}$ to $3\times10^{-3}$ after propagation through the sample. Thus we can conclude that the method used provides direct measurements of the magnitude and sign of the phase shift for the two linearly polarized components of light.

As shown in Eq. (17), in the case of low reflection and small thickness, Eq. (7) confines the phase difference to $|\tau| \approx 2\pi n' \Delta/\lambda$, so that $n' < 0$ results in $\tau < 0$. In experiments using interferometry, the phase shift due to a NIM layer can be precisely measured relative to a layer of air of the same thickness: $\tau_{\text{exp}} = \tau - \tau_{\text{air}}$, where the phase shift in air is a reference phase, $\tau_{\text{air}} = 2\pi\Delta/\lambda$. Then $n'$ is negative in the material provided that $\tau_{\text{exp}} < -\tau_{\text{air}}$. In general, for a NIM layer with reflection, one should also account for $\rho$ as in Eq. (18). In materials with strong

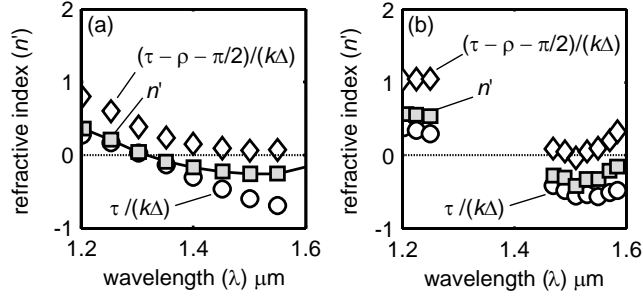

**Fig. 7. Refractive index of a NIM layer restored from FDTD simulations (a), and measurements (b). Refractive index n is found from exact formulas (11)-(13). Approximations obtained using Eqs. (17)-(18) are also shown for comparison**

absorption, the relation between complex parameters $t$, $r$ and $n$ is even more complicated and phase measurements should be accompanied by measurements of the transmittance and reflectance magnitudes in order to use the accurate procedure of Eqs. (11), (12) and (13).

To validate the procedure for the restoration of $n'$ and the phase-based approximations (Eqs. (17) and (18)), we use experimental data of our 3D NIM sample[21] (see below for more detailed consideration of this Sample B, consisting of pairs of coupled nanorods). The sample is arranged of a periodic array of coupled gold nanorods deposited directly on glass. Fig. 7 shows the refractive index obtained (a) from FDTD simulations, and (b) from measurements. A segment of the wavelength range where the refractive index becomes minimal is selected. The minimal value of $n'$ for this structure is about −0.3 at 1.5 µm. In contrast to the examples of Fig. 5, in this case approximations by Eq. (17) work better due to lower reflection.

## Numerical and physical experiments with NIMs

Up to now, we have seen that a negative refractive index is provided by resonant coupled metal-dielectric elementary scatterers. For example, our typical 2D models of Fig. 4(a) display negative refraction up to $n = -2$, as shown in Fig. 4(d) and Figs. 5(e) and (f). These strip structures can be readily fabricated, but losses are still rather large.

We now analyze the NIM-substrate interaction, beginning with the basic cases of a single gold rod on ITO-glass and pure glass substrates. We then consider several core 3D NIMs that demonstrate a negative refractive index proven in both experiments and simulations.[21-23]

### *An equivalent Debye model in FDTD*

Rather than utilizing the previously applied finite element method (FEM) to investigate the resonance behavior of 2D gold strips and to pinpoint a negative refractive index as an effective quantity, the numerical method we use for 3D structures involves the well known finite-difference time domain (FDTD) technique.[27] The FDTD modeling of plasmonic resonances in the optical range is more complicated than either a standard perfect electric conductor (PEC) approach for thin skin-depths or a direct application of conductivity.

The Drude model for a given single decay constant ($\Gamma$) and plasma frequency ($\omega_p$) is defined as,[28]

$$\varepsilon_r(\omega) = \varepsilon_\infty - \frac{\omega_p^2}{\omega(\omega + i\Gamma)}, \tag{19}$$

while the equivalent Debye model is defined as

$$\varepsilon_r(\omega) = \varepsilon_\infty + \frac{\chi_1}{1 - i\omega t_0} - \frac{\sigma}{i\omega\varepsilon_0}, \tag{20}$$

where $\varepsilon_\infty$ is the permittivity at infinite frequencies, $\chi_1 = \varepsilon_s - \varepsilon_\infty$ is the permittivity step, and $t_0 = 1/\Gamma$ is the relaxation time. Note that the Debye model can be straightforwardly derived from the Drude model (Eq. (19)) using a partial fraction expansion; this gives $\chi_1 = -(\omega_p t_0)^2$ and $\sigma = \omega_p^2 \varepsilon_0 t_0$.

Using Eq. (20) in $\vec{D} = \varepsilon(\omega)\vec{E}$, we obtain $\vec{D}(\omega) = \varepsilon_0 \left(\varepsilon_\infty \vec{E} + \vec{I}_1(\omega) + \vec{I}_2(\omega)\right)$, where

$$\vec{I}_1(\omega) = \frac{\chi_1}{1 - i\omega t_0} \vec{E}(\omega), \tag{21}$$

and

$$\vec{I}_2(\omega) = -\frac{\sigma}{i\omega\varepsilon_0} \vec{E}(\omega). \tag{22}$$

Separating each term in the time domain, we introduce a new displacement current vector $\vec{D}(t) = \varepsilon_0 \left(\varepsilon_r \vec{E} + \vec{I}_1(t) + \vec{I}_2(t)\right)$, where the time-domain versions of Eqs. (21) and (22) are defined by the following convolutions:

$$\vec{I}_1(t) = -\omega_p^2 t_0 \int_0^t e^{-(t-\tau)/t_0} \vec{E}(\tau) d\tau, \tag{23}$$

and

$$\vec{I}_2(t) = \omega_p^2 t_0 \int_0^t \vec{E}(\tau) d\tau. \tag{24}$$

Then, except for a new integration term ($\vec{I}_1(t)$), the calculation scheme is very similar to the standard eddy current problem.[27]

## Non-coupled nanorods on an ITO-glass or bare glass substrate

Consider, for example, FDTD modeling of the periodic metal-dielectric structure shown in Fig. 8(a) and (c). The elementary cell of the array is comprised of a single gold nanorod placed on an ITO-glass substrate. The restored values of the refractive index (using Eq. (16) and then Eqs. (11)–(13)) are shown in Fig. 9 for parallel (a) and perpendicular (b) polarizations of the incident light. Transmittance (T), reflectance (R) and absorbance (A) are also calculated for the same model with parallel and perpendicular polarizations (Figs. 9(c) and (d), respectively).

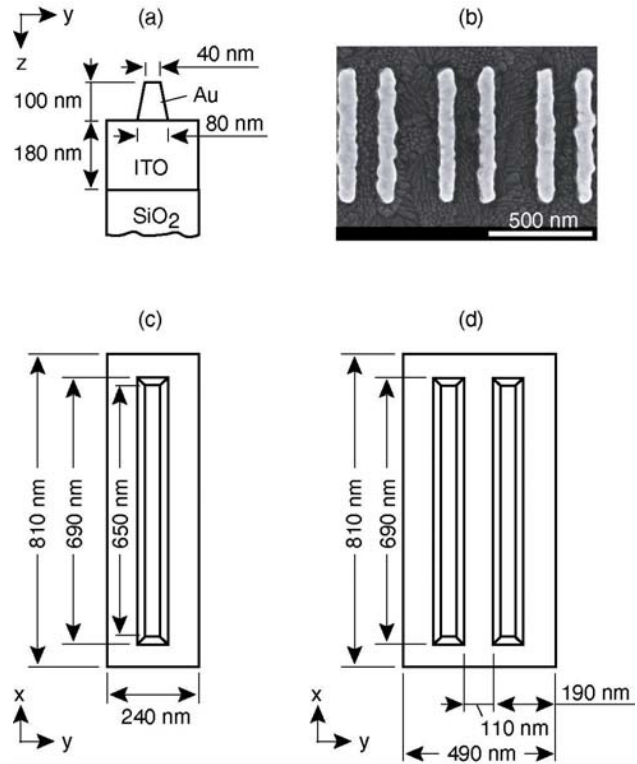

**Fig. 8. Single-periodic (a and c) and double-periodic (b and d) models for numerical simulations of non-coupled rods deposited on an ITO-glass or bare glass substrate. The geometry in (b) and (d) represents Sample A fabricated by electron-beam lithography.**

The basic single structure creates an almost zero equivalent refractive index,

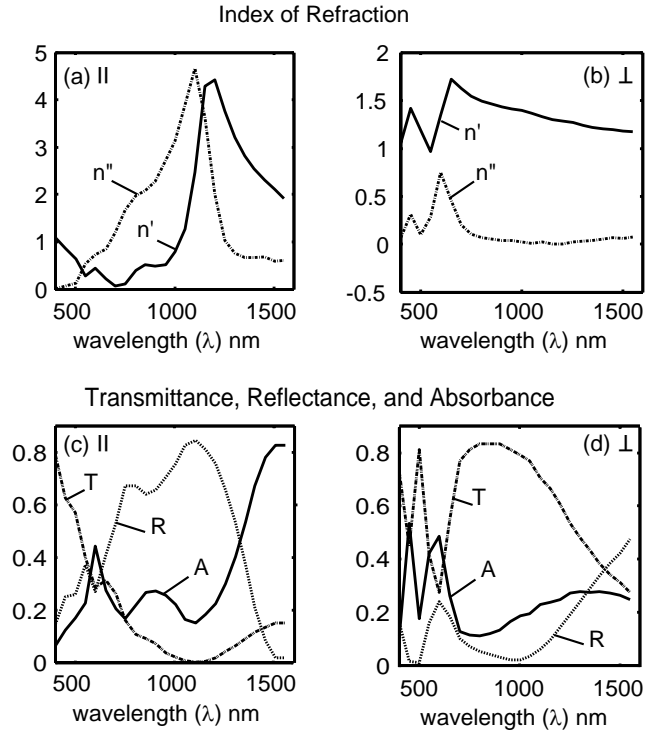

**Fig. 9. The index of refraction for parallel (a), and perpendicular (b) polarizations of incident light obtained from FDTD simulations with the single-periodic geometry of Fig. 8. Transmittance (T), reflectance (R) and absorbance (A) calculated for the same polarizations are shown in (c) and (d).**

demonstrates resonant absorbance for the parallel polarization, and displays a completely different behavior for the other polarization direction. (Following characteristic dimensions, resonant features for perpendicular polarizations are shifted toward much shorter wavelengths.)

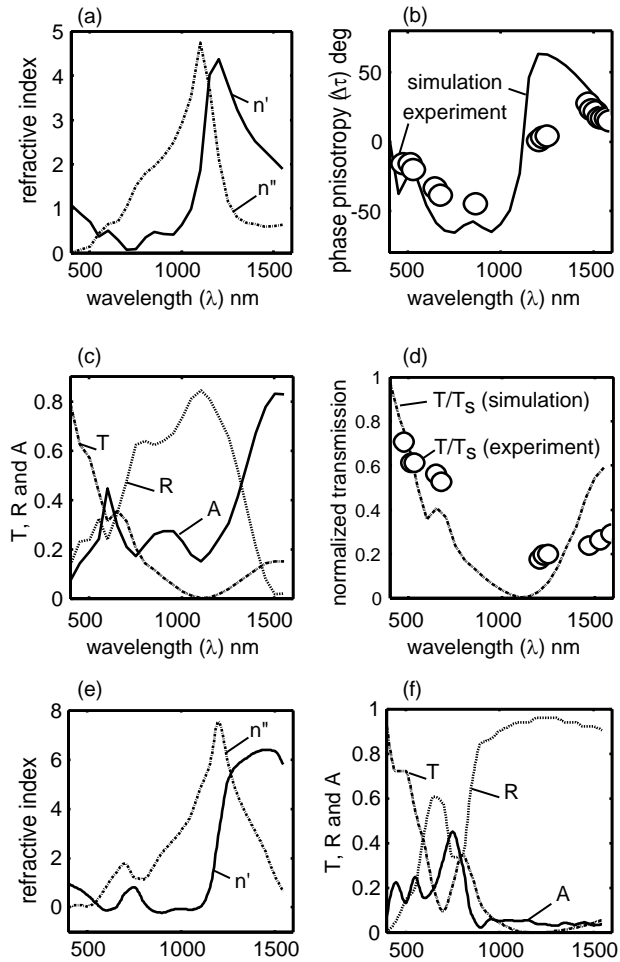

**Fig. 10. The index of refraction for parallel polarization (a) and the phase anisotropy in transmission (b) obtained from FDTD simulations with the double-periodic geometry of Fig. 8. Transmittance (T), reflectance (R) and absorbance (A) calculated for the same polarization are shown in (c). Calculated values of the normalized transmission ($T/T_s$) are compared to the experimental data in (d). Diagram (e) depicts the refractive index for the identical composite structure without any substrate. T, R, and A for this case are shown in (f).**

Upon obtaining the result, we suggest that for the transverse magnetic case, where the electric field is parallel to the periodic rods, we observe a coupled resonant behavior because, as shown below, the electric field is able to circulate in continuous contours at the rod-substrate interface. We also imply that strong coupling with a lossy substrate such as ITO could be unfavorable. For this reason, a novel double-periodic structure has been fabricated, tested and simulated using 3D FDTD code.

A scanning electron microscope (SEM) image of the new, double-periodic structure (called Sample A) is shown in Fig. 8(b), and a top view of the unit cell is shown in Fig. 8(d). The double-periodic structure consists of two rods which share the dimensions of the single rod in Fig. 8(a) and (c). We note that, in spite of more sophisticated periodicity, the double-periodic metallic elements exhibit very similar refractive index behavior and absorbance in both polarization directions. An example for the parallel polarization is shown in Figs. 10(a) and (c). That also suggests that *deviations (up to 20%) in periodicity do not change the equivalent optical properties of the periodic plasmonic structure*. We note that in spite of the non-idealities of Sample A, shown in Fig. 8(a), the phase anisotropy in Fig. 10(b) and the transmittance (T) normalized by the transmittance of the substrate ($T_s$) in Fig. 10(d) both satisfactorily match the experimental data.

As illustrated below, the rods are also coupled inductively to the ITO-glass substrate. The electric field component $E_z$, which is not mixed with the incident field, is mapped at *xy* cross-sections taken through the middle of the rod (see Fig. 11(a)) and just 10 nm beyond the rod inside the ITO (Fig. 11(b)). The rods are shown by dashed lines in Fig. 11. In both cases the two rods perform almost like a strip, forming a high-order evanescent mode inside ITO. In addition to *xy* cross-sectional field maps, Fig. 11 depicts the values of the electric field component $E_z$ in

(c) and magnetic filed component $H_y$ in (d) mapped at *xz* cross-sections through the middle of the rod. The values of $H_y$ are normalized by the incident magnetic field taken at the geometrical

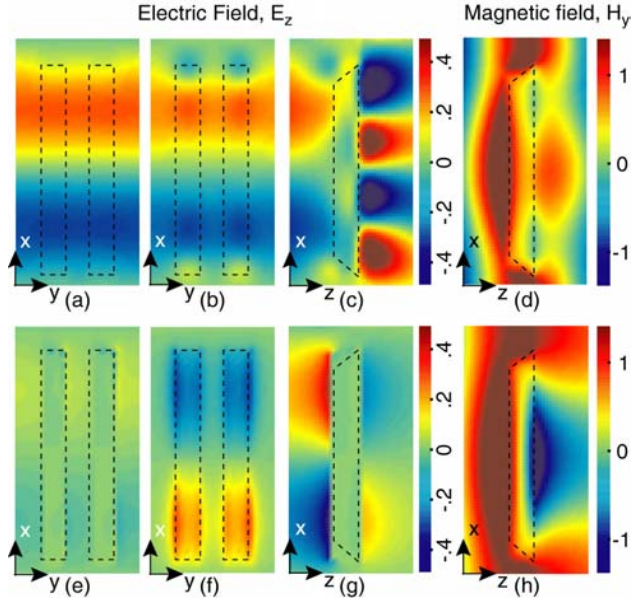

Fig. 11. Diagrams in top row, (a) through (d), depict field maps for non-coupled rods on ITO-glass substrate. Similar maps in (e) through (h) are shown for the sample without any substrate. In (a) and (e) the electric field component $E_z$ is mapped at xy cross-sections through the middle of the rod, while (b) and (f) are just 10 nm beyond the rod inside the ITO. $E_z$ in (c) and (g) and $H_y$ in (d) and (h) are mapped at xz cross-sections through the middle of the rod. All magnetic field values are normalized by the magnetic incident field taken at the geometrical center of the rod.

center of the rod.

The conductivity of ITO is large enough to allow for electric fields to circulate along continuous contours through the gold-ITO interface. After the incident magnetic field is added to the induced one, reversed magnetic field zones are observed right at the gold-ITO interface (see Fig. 11(d)).

Using FDTD analysis of non-coupled rods of Fig. 8(d), it was discovered that a simple substitution of the ITO-glass substrate with bare glass could provide a negative refractive index. Fig. 10(e) depicts FDTD results for the non-coupled rods of Fig. 8(d) but without any substrate. The minimal value of $n' \approx -0.2$ is achieved at a wavelength of 0.9 μm. The results are consistent with the more intense and larger field reversal zone shown in Fig. 11(h). We note a substantially different distribution of the electric component $E_z$ in this case. This observation of negative refractive index for single, non-coupled rods is in agreement with earlier theoretical predictions.[18]

*Sample B: Coupled-nanorods on a bare glass substrate*

In addition to results for single, planar, non-coupled rods, 3D coupled rods[17-21] with a magnetic field reversal similar to that above would be desirable. Such a NIM would consist of periodic arrays of coupled gold rods oriented parallel to the incident electric field and deposited directly on glass. An example unit cell is depicted in Fig. 12(a) and (c) and the results for the refractive index and transmittance, reflectance, and absorbance are shown in Figs. 13(a) and (b). The model demonstrates a refractive index of about −0.5 at 1360 μm, although with substantial losses around this point.

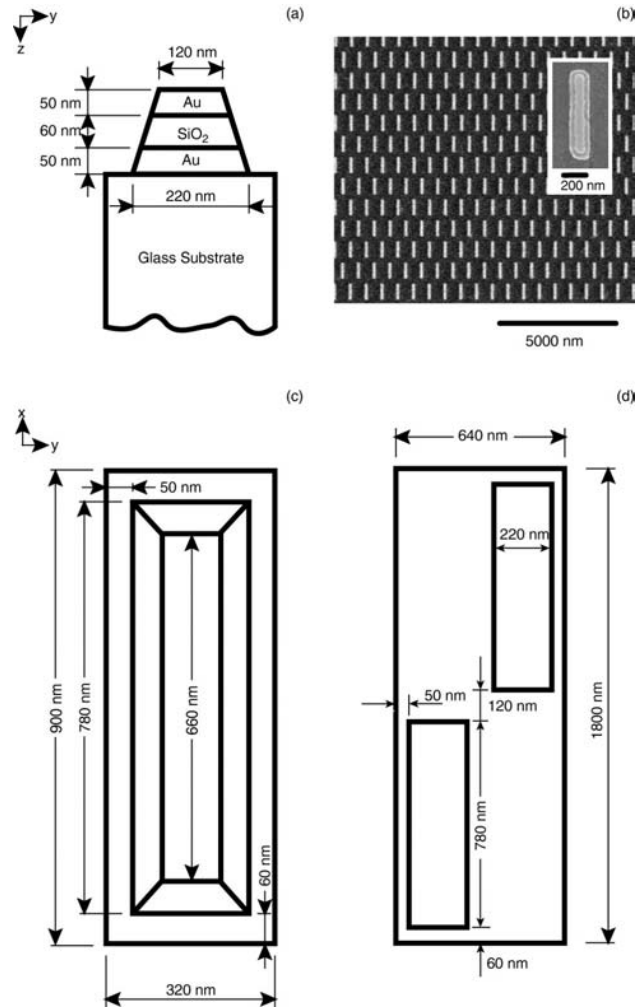

**Fig. 12. Single-periodic (c) and skewed-periodic (d) model structures of coupled gold rods deposited on a bare glass substrate. Both models used the same rod pairs shown in (a). The geometry in (d) simulates a fabricated sample (Sample B). SEM image of Sample B is shown in (b).**

To avoid a probable interaction between the rows of coupled rods, and to possibly decrease losses, an alternative sample (Sample B) with a skewed symmetry as shown in Fig. 12(b) and (d) has been manufactured and simulated. Sample B requires a unit cell four times larger than that considered for the structure of Fig, 12(a) and (c), and consequently one would expect a weaker negative refraction effect due to a smaller metal filling factor. Indeed, as shown in Fig. 14(a), the index of refraction obtained from FDTD simulations for parallel polarization is − 0.2 at 1.5 μm with the geometry of Fig. 12(b) and (d). Experimental studies of the NIM provide a value of $n = -0.3 \pm 0.1$ at 1.5 μm[21] (see also Fig. 7 above). Both the experimental and simulated values of the refractive index have already been compared to phase-only approximations in Fig. 7. Simulated and experimental transmittance ($T_{sim}$ and $T_{exp}$) and reflectance ($R_{sim}$ and $R_{exp}$) calculated for the same polarizations are shown in Fig. 14(c) and (d).

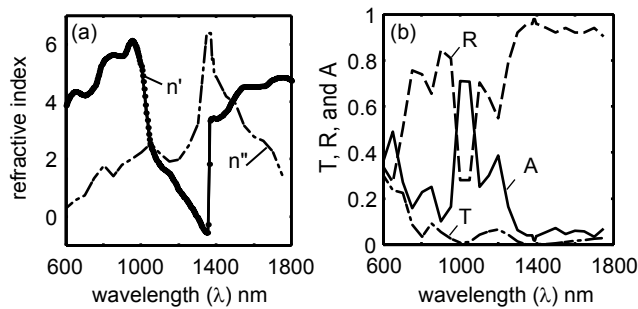

**Fig. 13. (a) The index of refraction for parallel polarization of incident light obtained from FDTD simulations with the single-periodic geometry of Fig. 12. (b) Simulated transmittance (T), reflectance (R), and absorbance (A) for parallel polarization.**

Direct deposition of the gold nanorod pairs on a bare glass substrate eliminates the damping effect from the ITO layer, and larger magnetic reactions are attainable due to this fabrication method. For testing our assumptions on the effect of an ITO-glass substrate on NIM properties, we have fabricated a preliminary sample on an ITO-glass substrate with a geometry very close to that of Sample B and compared it to the results from Sample B.

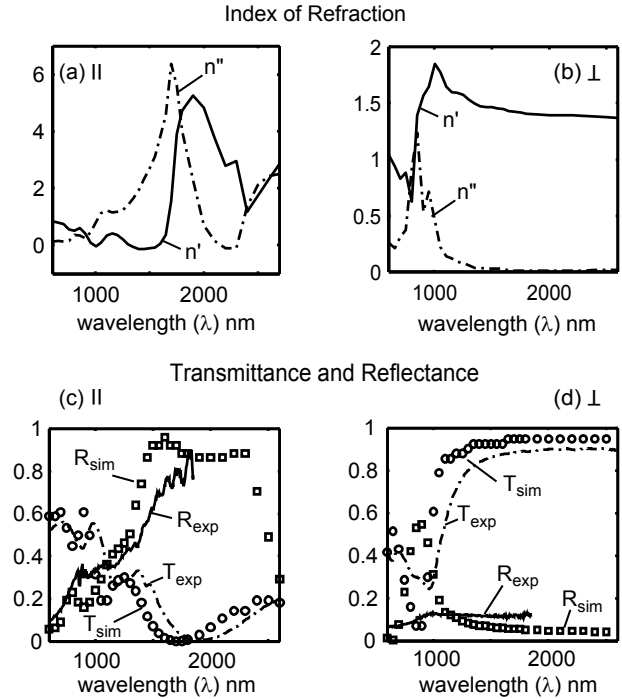

**Fig. 14. The index of refraction for parallel (a), and perpendicular (b) polarizations of incident light obtained from FDTD simulations with the double-periodic geometry of Fig. 8. Simulated and experimental transmittance ($T_{sim}$ and $T_{exp}$) and reflectance ($R_{sim}$ and $R_{exp}$) calculated for the same polarizations are shown in (c) and (d).**

First, in contrast to Sample B, the preliminary sample does not demonstrate a negative refractive index in either experiment or simulations. Second, the induced electric mode between the rods appears to be too high, and the magnitude of the induced magnetic dipole moment is too low for a good magnetic reaction in this case. That effect is illustrated in Fig. 15. The figure shows simulated maps of the electric field $E_z$ obtained for the two samples with similar geometries. Field maps in Fig. 15(a) and (b) represent the sample deposited on an ITO-glass substrate, while the maps in Fig. 15(c) and (d) are calculated for the same geometry deposited on bare glass (Sample B). $E_z$ is taken at two *xy* cross-sections, first through the middle of the rod, as shown in (a) and (c), and then at 10 nm beyond the rod inside the substrate, as depicted in (b) and (d). Comparison of (a) to (c) allows us to conclude that the elimination of ITO produces stronger currents in the required mode between the rods. Weaker higher-order current modes are generated between the rods in the sample with an ITO-glass substrate, as shown in Fig. 15(a).

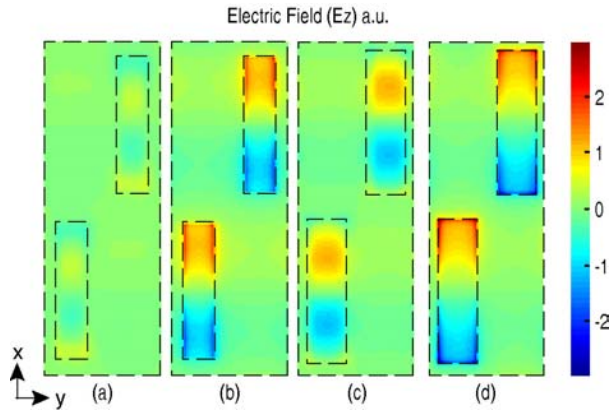

**Fig. 15. Simulated maps of $E_z$ obtained for two samples. Field maps (a) and (b) are calculated for the sample deposited on an ITO-glass substrate, while (c) and (d) are calculated for a similar sample deposited on bare glass (Sample B). Electric field ($E_z$) is mapped at two xy cross-sections: through the middle of the pair of rods, as in (a) and (c), or just 10 nm beyond the rod inside ITO or glass, as in (b) and (d).**

In addition, both samples are not optimized for the best magnetic reaction, since the second rod in both cases does not "channel" a sufficient part of the electric field. Indeed, we have assumed that $E_z$ should be at least weaker (or reversed) just beyond the second rod. Unfortunately in both samples, as indicated in Fig. 15(b) and (d), $E_z$ neither decays nor changes its direction at the cross-section just behind the rods. In essence, the presence of any substrate alters the symmetry of the interacting rods and restrains the magnetic reaction between the rod pair. This defect could be alleviated by optimizing the design of the structure.

## *Inverted nanorods: coupled elliptic voids in metallic films*

To achieve a NIM within a given range of optical wavelengths, an inverted structure could also be employed along with dielectric spacers between coupled metallic elements. Such a design offers a good manufacturability for a thick multilayer NIM.

An FDTD model is created to match the experimental sample of Ref. [23]; the dimensions used in simulations are rounded (shown in parentheses). The basis of the structure consists of a three-layer film composite deposited over a BK7 glass substrate. The composite consists of two 30 (30) nm layers of gold separated by a 75 (80) nm layer of $Al_2O_3$ as shown in Fig. 16(a). An array of elliptical wells (voids) is then etched out of the three layer composite to give the final sample with a 2D lattice period of 787 (790) nm. The dimensions of the elementary cell with the well are depicted in Fig. 16(a) and (b). A mid-section view of the elementary cell along the short axis of the elliptic well is shown in Fig. 16(a).

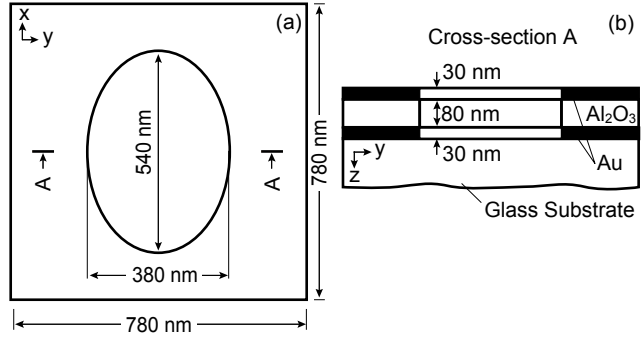

**Fig. 16. (a) An elementary cell of coupled elliptic voids as an inversion of coupled nanorods. (b) A cross-sectional view of the elementary cell shown in (a).**

Fig. 17 shows the simulated index of refraction for parallel (a) and perpendicular (b) polarizations of incident light for the geometry of Fig. 16 together with phase-based approximations by Eqs. (17) and (18). Transmittance ($T$), reflectance ($R$) and absorbance ($A$) are shown in (c) and (d). As expected, a simple approximation with Eq. (17) works well in the case of moderate reflection (see Fig. 17(a) and (c)) in comparison with the case of large $R$ (see Fig. 17(b) and (d)). The other approximation, Eq. (18), provides a better fit in all cases.

The inverted NIM demonstrates a negative refractive index of −1.5 at 1.8 μm for perpendicular polarization (electric field is along the short axis of the elliptic well) and −0.75 at 1.6 μm for parallel polarization (electric field is along the long axis of the elliptic well), which is consistent with the experimental data[23]. As expected for inverted structures, the perpendicular polarization works better than the parallel one.

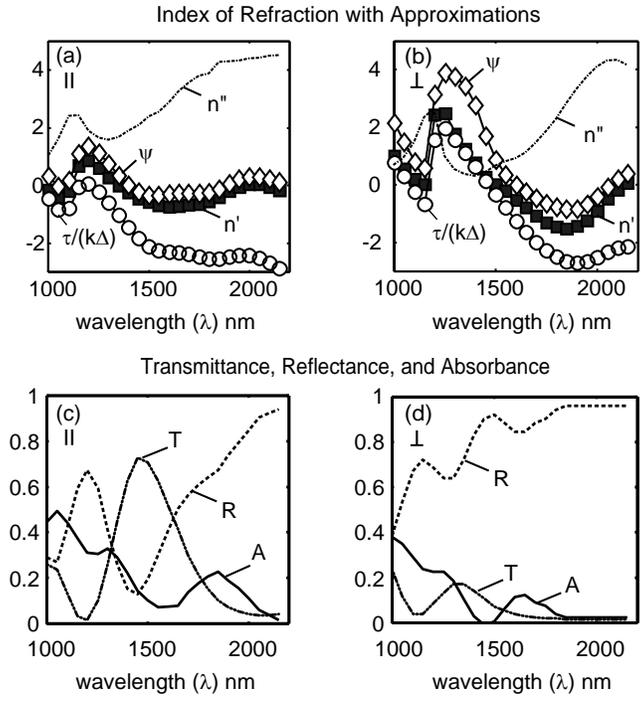

**Fig. 17. The index of refraction for parallel (a) and perpendicular (b) polarizations of incident light obtained from simulations for the inverted geometry of Fig. 16. Approximations with Eqs. (17) and (18) are also tested for the inverted geometry. Transmittance (T), reflectance (R) and absorbance (A) calculated for the same polarizations are shown in (c) and (d).**

## Summary

We have described and discussed several thin composite metal-dielectric structures which demonstrate an effective negative refractive index. We considered examples of how the complex values of the transmission and reflection coefficients ($t$ and $r$) taken either from measurements or simulations could be used to obtain the index of refraction. From $t$ and $r$ for the incident plane wave, we calculated the effective refractive index for all of the composite materials considered herein. The description is based on an effective layer of homogeneous media with the same effect on $t$ and $r$ as a given layer of a NIM of the same thickness. Characteristic matrices are used to obtain the effective impedance and the effective refractive index for a single unknown NIM layer in an arbitrary multilayer structure.

Our simulations show that a composite NIM based on a 2D periodic array of coupled gold strips separated by a continuous dielectric layer can have a negative refractive index of down to −2.

We also showed the critical importance of phases for retrieval of the refractive index. We have found that formulas using only the phases of $t$ and $r$ could represent a good alternative to exact restoration of the refractive index. Consequently, we propose two interferometry schemes, the polarization interferometer for measuring phase anisotropy between two different polarizations in transmission (reflection), and the walk-off interferometer for measuring absolute phases.

We found that non-coupled gold rods arranged on an ITO-glass substrate do not provide a negative refractive index because of the damping associated with the conductive features of ITO. We deduce that losses in ITO introduce a barrier to achieving negative refraction with ITO-glass substrates. Simulations indicate that an identical structure on a glass substrate could give $n' \approx -0.2$.

Experimental studies for a sample based on an array of paired metal rods on a glass substrate allowed us to observe a negative refractive index of −0.3 at 1.5 μm, in good agreement with our simulations.

Finally, we showed that inverse structures also provide good results in creating NIMs. We analyzed an inversion of the coupled rods structure, where the coupled gold rods are substituted by coupled voids in gold films. Validating the design suggested in Ref. [23], we obtained a refractive index of −1.5 at 1.8 μm for perpendicular polarization (the electric field along the short axis of the elliptic well) and −0.75 at 1.6 μm for parallel polarization. Again using our suggested phase-based approximations, we found that, similar to non-inverted designs, only the phases of $t$ and $r$ are sufficient to provide estimates of the negative refraction effect in inverted designs with voids.

We expect that further optimization of design techniques for composite NIMs will provide a stronger effect at smaller losses, enabling new devices based on scalable 3D NIM structures.

This work was supported in part by ARO grant W911NF-04-1-0350 and by NSF-NIRT award ECS-0210445.
References:

1. L. I. Mandel'shtam, "Group velocity in crystal lattice," JETP **15**, 475 (1945); also in, L. I. Mandel'shtam, "The 4th Lecture of L.I. Mandel'shtam given at Moscow State University (05/05/1944)," Collection of Scientific Works, Vol. 2 (Nauka, Moscow, 1994).

2. V. G. Veselago, "The electrodynamics of substances with simultaneously negative values of $\varepsilon$ and $\mu$," Soviet Physics Uspekhi **10**, 4, 509-514 (1968).

3. H. Lamb, "On group-velocity," Proc. London Math. Soc., Ser. 2, **1**, pp. 473-479, 1904.

4. A. Schuster, *An Introduction to the Theory of Optics* (Edward Arnold, London, 1904).

5. J. B. Pendry, "Negative refraction makes a perfect lens," Phys. Rev. Lett. **85**, 3966-3969 (2000).

6. V. A. Podolskiy, E. E. Narimanov, "Near-sighted superlens," Optics Letters **30**, 75-77 (2005)

7. K. J. Webb, M. Yang, D. W. Ward, and K. A. Nelson, "Metrics for negative-refractive-index materials," Phys. Rev. E **70**, 035602(R), (2004).

8. N. A. Khizhnyak, "Artificial anisotropic dielectrics formed from two-dimensional lattices of infinite bars and rods," Sov. Phys. Tech. Phys. **29**, 604-614 (1959).

9. A. N. Lagarkov and A. K. Sarychev, "Electromagnetic properties of composites containing elongated conducting inclusions," Phys. Rev. B **53**, 006318 (1996).

10. T. J. Yen, W. J. Padilla, N. Fang, D. C. Vier, D. R. Smith, J. B. Pendry, D. N. Basov, and X. Zhang, "Terahertz magnetic response from artificial materials," Science **303**, 1494-1496 (2004).

11. S. Linden, C. Enkrich, M. Wegener, J. Zhou, T. Koschny, and C. Soukoulis, "Magnetic response of metamaterials at 100 terahertz," Science **306**, 1351-1353 (2004).

12. S. Zhang, W. Fan, B. K. Minhas, A. Frauenglass, K. J. Malloy, and S. R. J. Brueck, "Midinfrared resonant magnetic nanostructures exhibiting a negative permeability," Phys. Rev. Lett. **94**, 037402 (2005).

13. A.K. Sarychev and V.M. Shalaev, "Magnetic Resonance in metal nanoantennas," in *Complex Mediums V*, Proc. SPIE **5508**, 128-137 (2004).

14. C. Enkrich, M. Wegener, S. Linden, S. Burger, L. Zschiedrich, F. Schmidt, J. Zhou, Th. Koschny, and C.M. Soukoulis, "Magnetic metamaterials at telecommunication and visible frequencies," arXiv:cond-mat/0504774v1, 29 Apr. 2005

15. A. Berrier, M. Mulot, M. Swillo, M. Qiu, L. Thylén, A. Talneau, and S. Anand, "Negative Refraction at Infrared Wavelengths in a Two-Dimensional Photonic Crystal," Phys. Rev. Lett. **93**, 073902 (2004).

16. E. Schonbrun, M. Tinker, W. Park and J.-B. Lee, "Negative Refraction in a Si-Polymer Photonic Crystal Membrane", IEEE Photon. Technol. Lett. **17**, 1196-1198 (2005).

17. V. A. Podolskiy, A. K. Sarychev, and V. M. Shalaev, "Plasmon modes in metal nanowires and left-handed materials," J. of Nonlinear Opt. Physics and Materials **11**, 65-74 (2002).

18. V. A. Podolskiy, A. K. Sarychev, and V. M. Shalaev, "Plasmon modes and negative refraction in metal nanowire composites," Opt. Express **11**, 735-745 (2003).

19. V. A. Podolskiy, A. K. Sarychev, E. E. Narimanov and V. M. Shalaev "Resonant light interaction with plasmonic nanowire systems," J. Opt. A: Pure Appl. Opt. **7**, S32–S37 (2005).

20. A. K. Sarychev, V. P. Drachev, H.-K. Yuan, V. A. Podolskiy, and V. M. Shalaev, "Optical properties of metal nanowires," SPIE Proceedings, **1**, 5219, San Diego (2003).

21. V. M. Shalaev, W. Cai, U. Chettiar, H.-K. Yuan, A. K. Sarychev, V. P. Drachev, and A. V. Kildishev, "Negative Index of Refraction in Optical Metamaterials," arXiv:physics/0504091, http://arxiv.org/ftp/physics/papers/0504/0504091.pdf (Apr. 13, 2005).

22. S. Zhang, W. Fan, N. C. Panoiu, K. J. Malloy, R. M. Osgood, S. R. J. Brueck, "Demonstration of near-infrared negative-index materials," http://arxiv.org/ftp/physics/papers/0504/0504208.pdf (2005).

23. S. Zhang, W. Fan, K. J. Malloy S. R. J. Brueck, N. C. Panoiu and R. M. Osgood, "Demonstration of metal-dielectric negative-index metamaterials with improved performance at optical frequencies," (current issue of JOSA B).

24. J. D. Jackson, *Classical Electrodynamics*, (Willey, New York, 1962), pp. 291.

25. Y. Svirko, N. Zheludev and M. Osipov "Layered chiral metallic microstructures with inductive coupling", Appl. Phys. Let., **78,** 498-500 (2001).

26. D. R. Smith, S. Schultz, P. Markoŝ, and C. M. Soukoulis, "Determination of effective permittivity and permeability of metamaterials from reflection and transmission coefficients," Phys. Rev. B **65**, 195104 (2002).

27. A. Taflove and S. Hagness, *Computational Electrodynamics: The Finite-Difference Time-Domain Method* (Artech, Boston, 2000).

28. A. Ishimaru, *Electromagnetic Wave Propagation, Radiation, and Scattering* (Prentice Hall, 1991).